\documentclass[useAMS,usenatbib]{mn2e}
\usepackage{graphicx}

\newcommand\ion[2]   {#1\,{\sc #2}}
\newcommand {\nii}   {[\ion{N}{ii}]}
\newcommand {\oiii}  {[\ion{O}{iii}]}
\newcommand {\oi}    {[\ion{O}{i}]}

\newcommand {\ha}    {\ifmmode ${H}$\alpha \else {H}$\alpha$\fi}
\newcommand {\hb}    {\ifmmode ${H}$\beta  \else {H}$\beta$\fi}
\newcommand {\hg}    {\ifmmode ${H}$\gamma \else {H}$\gamma$\fi}
\newcommand {\laa}   {\ifmmode ${Ly}$\alpha \else {Ly}$\alpha$\fi}
\newcommand {\ergs}  {\ifmmode ${\ergs}$ \else ergs\,cm$^{-2}$\,s$^{-1}$\fi}
\newcommand {\ergsA} {\ergs\,\ifmmode ${\AA}$^{-1} \else \AA$^{-1}$\fi}
\title[Spectral variability of 3C~390.3]{Spectral variability of the 3C~390.3 nucleus for more than twenty years. I. Variability of the broad and narrow emission-line fluxes}
\author[S. G. Sergeev et al.]{S.~G.~Sergeev,\thanks{E-mail:
sergeev.crao@mail.ru} S.~V.~Nazarov, G.~A.~Borman
\\
Crimean Astrophysical Observatory, P/O Nauchny Crimea 298409, Russia}

\begin{document}

%\date{Accepted 1988 December 15. Received 1988 December 14; in original form 1988 October 11}

\pagerange{\pageref{firstpage}--\pageref{lastpage}} \pubyear{}

\maketitle

\label{firstpage}

\begin{abstract}
We summarize results of the analysis of the optical variability
of the continuum and emission-line fluxes in the 3C~390.3 nucleus
during 1992--2014. The \oiii\,$\lambda5007$ flux increases monotonically by $\approx 30$ per cent
in 2003--2014. The narrow Balmer lines show similar monotonic
increase, while the variability patterns of the \oi\,$\lambda6300$ narrow line are
completely different from that of \oiii.
The reverberation lags are found to be $88.6\pm8.4$, $161\pm15$,
and $113\pm14$\,d for the \hb, \ha, and \hg\ broad emission-lines, respectively.
The reverberation mass of the central black hole
equals to $(1.87\pm0.26)\times10^9\,M_\odot$ and
$(2.81\pm0.38)\times10^9\,M_\odot$, for the \hb\ and \ha\ lines
and assuming a scaling factor that converts the virial product to a mass to be $f=5.5$.
A difference between both masses can point to
a difference between kinematics of the \ha\ and \hb\ emission
regions.
We show that the reverberation mapping can only be applied to the entire
period of observations of the 3C~390.3 nucleus after
removing a long-term trend. This trend has been expressed by a slowly varying scale
factor $c(t)$ in the power-law relationship
between the line and continuum fluxes:  $F_{line}\propto c(t)\,F_{cont}^a$.
We find the power-law index $a$ equals to $0.77$ and $0.54$ for the \hb\ and \ha\ lines,
respectively.
The observed relationship between the Balmer decrement
and the optical continuum flux is as follows:
$F(\ha)/F(\hb) \propto F_{cont}^{-0.20}$ and
$F(\hb)/F(\hg) \propto F_{cont}^{-0.18}$.
The 3C~390.3 nucleus is an `outsider'
in the relationship between optical luminosity
and black hole mass. Its Eddington ratio is $E_{bol}/E_{Edd} = 0.0037$.
\end{abstract}


\section[]{Introduction}

Photoionization by central source radiation in active galactic nuclei (AGN)
is believed to be a main mechanism responsible for the gas emission
in the broad-line region (BLR).
So, the
observed correlations between broad-line properties and continuum properties provides
information necessary to develop photoionization models of the BLR.
These models together with the observed line intensities and line intensity ratios are used to recover physical
conditions in the BLR. The size, geometry, and
kinematics of the BLR can be determined
from the line-continuum correlation
by the reverberation mapping technique \citep{bla82, pet93, pet94}.
The most obvious success in the reverberation mapping studies
is determination the BLR size
for a lot of AGNs via the cross-correlation function. This function is applied to
the light curves of the continuum and emission lines to find
delays between variations of their fluxes \citep[e.g.,][]{Pet98b,Kas00,Ben10,Den10,Gri12}.
Such studies require a good sampling of observations as well as
significant variability amplitude with at least one event clearly
seen in the continuum light curve.
With the known BLR sizes and under some assumptions (such as virialized
gas motion in the BLR), it is
possible to determine the fundamental AGN characteristics --
the central black hole mass and accretion rate \citep[e.g.,][]{pet04, ben09}.

Recent years, good progress has been achieved in applying the
reverberation method to individual profile segments of the broad emission lines
\citep[e.g.,][]{den09,Gri12,Dor12}.
These reverberation results suggest that the BLR kinematics
can be distinctly different (inflow, outflow, or virialized motions) for
individual AGNs.

The line emission from the narrow-line region (NLR) is believed to be constant
in flux during several decades because this region is very extended spatially
(hundreds of light years).
However, some authors have claimed the narrow-line variability for a shorter period.
For example, \citet{Pro92} have carefully assumed that
``Time scale variability in most cases is several
years, but months and weeks should not be excluded''
for the \oiii\,$\lambda\lambda\,5007,4959$\,\AA\ forbidden lines
in some nuclei.
\citet{Pet13} have found that
``The narrow \oiii\,$\lambda\lambda\,5007,4959$ emission-line fluxes in the spectrum of the well-studied Seyfert 1 galaxy NGC~5548
are shown to vary with time.'' and they suggested the size of the \oiii-emitting
region in NGC~5548 to be 1--3\,pc.
\citet{Zhe95} have studied the intensities of the narrow lines in
3C~390.3 between 1974 and 1990. They have found
the flux of \oiii\,$\lambda4959$ has declined by 40 per cent between 1974 and 1984. It
has remained at a low level between 1984 and 1987, and then increased to
a higher level since 1988, in step with the long-term variation of the
underlying optical continuum.
In most papers, the light curves of the continuum and broad emission lines
are calculated using narrow line intensities as `photometric standards'.
Therefore, narrow line variability over
a period of observations leads to the appearance of fictive longtime `trends' in the
derived light curves.

The 3C~390.3 nucleus is often considered as a prototype for a relatively small group of AGNs
(referred to as `double-peaked emitters') with broad and double-peaked
low-ionization lines in their spectra.
The Balmer emission lines in the optical spectrum of the 3C~390.3 nucleus
are very broad with strong blueshifted and redshifted peaks
that have been well known long time ago \citep[e.g.,][]{san66, lyn68, bur71}.
It is one of the best studied nucleus among double-peaked emitters
\citep*[e.g.,][]{ost76, bar80, yee81,
net82, vei91, zh96, wam97, die98, obr98, Sha01, Ser02, gez07, Sha10, Ser11, Die12, Afa15}

Among various BLR models, the
disc-like BLR around the central black hole is widely accepted
to explain the broad emission-line profiles in double-peaked emitters.
Double-peaked broad lines being well fitted
by the emission from the relativistic circular accretion disc
within a distance from the black hole of hundreds to thousands of gravitational
radii \citep*{ch89a, ch89b}.
Modifications of this model include an elliptical accretion disc model
\citep{Era95}, a stochastically perturbed accretion disc model
\citep{Flo08},  a system of
clouds rotating predominantly in the same plane \citep{Ser00}, etc.
However, the BLR geometry in 3C3~90.3 seems to be very complex with
at least two components,
although the disc-like component is, probably, the dominant
emitter \citep{Ser02,Sha10,Afa15}.

There is some inconsistency among published lag measurements in 3C~390.3.
The lag between the continuum and Balmer
emission-line variations was found to be $\approx$90--180 days from the
long-term (several years) monitoring campaigns
\citep{Sha01,Ser02,Sha10,Ser11,Afa15}, more than twice the lag found
from the short-term campaigns \citep{die98,Die12}.
A summary of the lag estimates is given in \citet{Kov14}.

The present study is the continuation of the previous studies of 3C~390.3
nucleus carried out in the Crimean Astrophysical Observatory (CrAO)
since 1992 \citep{Ser02,Ser11}.
Now we present new results of the optical spectroscopic observations
of 3C~390.3 during monitoring campaign from 2008 to 2014 and we summarize
the results obtained on 3C~390.3 in 1992--2014.
The main goals of the present paper are similar to that of \citet{Ser11}:
(1) To determine the BLR size and the black hole mass via reverberation
mapping technique; (2) To check for the narrow-line variability; and
(3) To determine power-law indices of the relationship between
Balmer line and continuum fluxes and to look for any changes in the
Balmer decrement.
The structure of the present paper is similar to the paper of
\citet{Ser11}.

We analyze the
total line fluxes only, not the line profiles.
In the next papers of this series
we plan to study variations of individual profile segments of the broad
Balmer lines, including the 2-d reverberations mapping
and the long-term evolution of the broad emission-line profiles.
We hope our results will be used to critically
test competing models of the double-peaked emitters.

\section[]{Observations and data reduction}
\subsection[]{Optical spectroscopy}

Optical spectra of 3C~390.3 have been obtained at the 2.6-m Shajn telescope
of the CrAO since 1992. The results of the observations for the periods
1992--2000 and 2001--2007 are published in \citet{Ser02,Ser11}.
The spectra were registered at the two
separate spectral regions centered at the H$\alpha$ and H$\beta$ lines.
Since July 2005 the old Astro-550 CCD was replaced by a more modern SPEC-10 CCD.
More details about observations, observational setup,
and spectral data processing are in \citet{Ser02,Ser11}.

Our new data set in 2008--2014 consists of 96 spectra in the H$\beta$ region
and 30 spectra in the H$\alpha$ region.
The signal-to-noise ratio
(S/N) per pixel at the continuum level
is in the range 21--109 (mean value is 51) for the H$\beta$ region
and in the range 31--79 (mean value is 50) for the H$\alpha$ region.

The spectra of the 3C~390.3 nucleus have been calibrated in flux as described in
\citet{Ser02}. The final step of this calibration is the scaling of spectra to match the
fluxes of the selected narrow emission lines which are assumed to be constant
over time-scales of the monitoring programme. This assumption is
justified by the large spatial extent of the narrow line region (NLR). In the
case of 3C~390.3 the underlying difficulty is that there is some evidence for
narrow-line variability \citep{cl87, Zhe95, Ser11}. We discuss this problem in section
\ref{phot}.

We chose the narrow \oiii\,$\lambda$5007 (H$\beta$ region) and
\oi\,$\lambda$6300 (H$\alpha$ region) emission lines as internal flux standards.
Their absolute fluxes were measured from the spectra obtained under photometric
conditions by using the spectra of the comparison star \citep[see][]{Ser02}.

As in \citet{Ser02,Ser11} the line fluxes were measured
by integrating the flux above the underlying continuum within the selected
wavelength intervals.
The continuum fluxes were determined at ~5100\AA\, and ~6200\AA\,
in the rest frame of 3C~390.3, and designated them as $F_{5100}$ and $F_{6200}$,
respectively.

The uncertainties in our H$\alpha$-region fluxes are much greater than in the
H$\beta$-region ones because they are dominated by the uncertainties
in the flux-scaling
factors determined from the relatively weak \oi\,$\lambda$6300 narrow line.

\subsection[]{Optical photometry}
\label{phot}

Regular CCD photometric observations of the selected AGNs have been
started at the CrAO in 2001.
The instrumentation, reductions and measurements of our photometric data are
described in \citet{dor05} and \citet{se05}.

\begin{figure}
  \includegraphics[width=84mm]{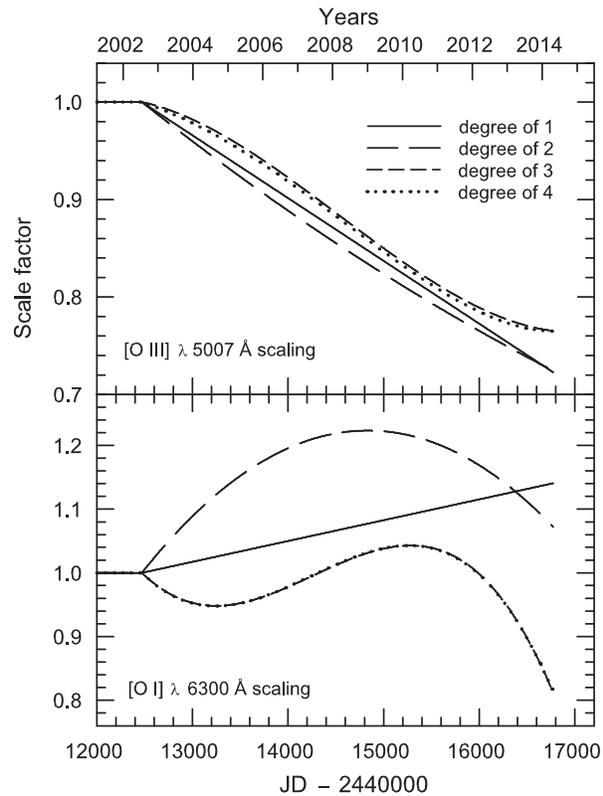}
  \caption{Normalized scale factor $\varphi$ as a polynomial function of time to scale our {\it V\/} filter
photometry to match the $F_{5100}$ (top panel) and $F_{6200}$ (bottom panel)
continuum fluxes. The legend for polynomial degrees is shown in the top panel.
The $\varphi$ is adopted to be 1 before JD2452472.5.}
  \label{scale}
\end{figure}

Our {\it V\/} filter photometric measurements of
3C~390.3 were calibrated to match the $F_{5100}$ continuum fluxes
as described in \citet{Ser11}, namely:

\begin{equation}
\label{eq-calib}
F_{5100} = \varphi\,(10^{-0.4V} - G)\,,
\end{equation}

where {\it V\/} is the stellar magnitude corrected for the broad H$\beta$ line,
$\varphi$ is a scale factor, and $G$ is a difference in constant contributions
(host galaxy and NLR).
The parameters $\varphi$ and $G$ were determined by the ordinary linear
regression.
Just the same as in \citet{Ser11} we have found strong discrepancy
in our spectral and photometric measurements that can only be removed by
assuming long-term changes in the scale factor $\varphi$. Now we have enough
absolute measurements of the \oiii\,$\lambda\,5007$\,\AA\ line fluxes made
using the comparison star during nights with good weather conditions to prove that
above changes must be attributed to the narrow-line variability.
We present the results on this variability in section \ref{narvar}.
So, in contrast to \cite{Ser11}, we assumed that our photometric
measurements are correct, while our spectral measurements are wrong
because variability of the narrow lines. To correct our spectral measurement
we recovered $\varphi(t)$ function as described in \citet{Ser11}.
We selected it to be a small degree polynomial function of time.
The polynomial coefficients and offset $G$ were determined
by achieving the best rms agreement between photometric and spectral data sets.
We adopted the $\varphi(t)$ function to be constant ($\varphi=\varphi_0$)
before beginning of the systematic continuum brightening in $\approx$2003.
Total of 105 data points were used for the fit.
The best fit normalized function $\varphi(t)/\varphi_0$
is shown in Fig.~\ref{scale}. It was computed for the degrees of
polynomial function from 1 to 4. The $\varphi(t)$ function is very similar
for all polynomial degrees and it increases monotonically by $\approx 30$ per cent
during about ten years.
The best solution for the polynomial degree 3 is as follows:

\begin{equation}
\label{eq-phi}
\frac{\varphi(t)}{\varphi_0}=\left\{
\begin{array}{l}
0.8790 - 0.07324\,\Delta t + 0.002255\,\Delta t^2 +\\
0.004117\,\Delta t^3,       for~ t>2452472.5 \\
\\
1,                          for~ t\le 2452472.5 \\
\end{array}
\right.
\end{equation}

where $\Delta t$ is adopted to be $\frac{t-2454713.1}{1000}$, $t$ is Julian Date,
and $\varphi/\varphi_0$ is adopted to be 1 before
JD2452472.5, i.e., before beginning of the systematic continuum brightening.
The time of $2454713.1$ is simply the averaged value of $t$, so we adopt
$\Delta t = 0$ for the mean time $t$.

However, the same method applied to the \ha-region spectra (scaled
using \oi\,$\lambda$6300\AA\ narrow line) gives ambiguous result
(see Fig.~\ref{scale}, bottom panel) where the recovered $\varphi(t)$ functions
are too different for various polynomial degrees.
Absolute flux measurements of the \oi\,$\lambda$6300\AA\ line (see section \ref{narvar})
show different variability patterns and smaller variability amplitude
as compared to the \oiii\,$\lambda\,5007$\,\AA. Because our results on the \oi\
variability as well as on the $\varphi(t)$ function recovery
are too unreliable, our \ha-region fluxes
were not corrected for the narrow-line variability.

The reliability of the $\varphi(t)$ fit for the \oiii\,$\lambda$5007 line (as well
as its unreliability for the \oi\,$\lambda$6300 line) is clearly seen
from Table~\ref{phirms} where fractional rms agreement
(corrected for degree-of-freedom)
between spectral and photometric data sets are shown.
In particular, the assumptions that the $\varphi(t)$ for the \oiii\,$\lambda$5007
is a straight line improves the rms agreement by 3 times, while the same assumption
for the \oi\,$\lambda$6300 line does not improves it any
significantly.

\begin{table}
\caption{Quality of the $\varphi(t)$ function fit given as a fractional rms agreement (in per cents)
between spectral and photometric data sets}
\label{phirms}
\begin{tabular}{lccccc}
\hline
Calibration  & \multicolumn{5}{c}{rms agreement for polynomial degree of} \\
line         &  0   &   1   &    2    &    3    &    4\\
\hline
\hline
\oiii\,$\lambda$5007    &  6.984  & 2.375 &  2.338  & 2.286 & 2.285\\
\oi\,$\lambda$6300      &  5.277  & 5.258 &  4.875  & 4.445 & 4.449\\
\end{tabular}
\end{table}

\subsection{Final light curves}

As mentioned in the previous section, we found that
our photometric $V$-band measurements are correct,
while our continuum light curve $F_{5100}$ has a long-term trend
because long-term flux changes in the \oiii\,$\lambda$5007 line used
for calibration of the \hb-region spectra. To merge the photometric and spectral data sets
we have first put the photometric light curve into the scale of the
continuum spectral fluxes (\ergsA) by subtracting an offset $G$ and by multiplying
by a factor $\varphi=\varphi_0$.
Then the spectral fluxes ($F_{5100}$, \hb, and \hg)
were corrected for the narrow-line variability
by dividing them by $\varphi(t)/\varphi_0$ (see eq.~\ref{eq-phi}).
Because of this correction, some values quoted
in the present paper may differ slightly from those in \citet{Ser11}.
The joined continuum light curve has been nightly averaged.

We do not perform any corrections for the \ha-region fluxes because
our results on the \oi\,$\lambda$6300 variability are too unclear.

\begin{figure}
  \includegraphics[width=84mm]{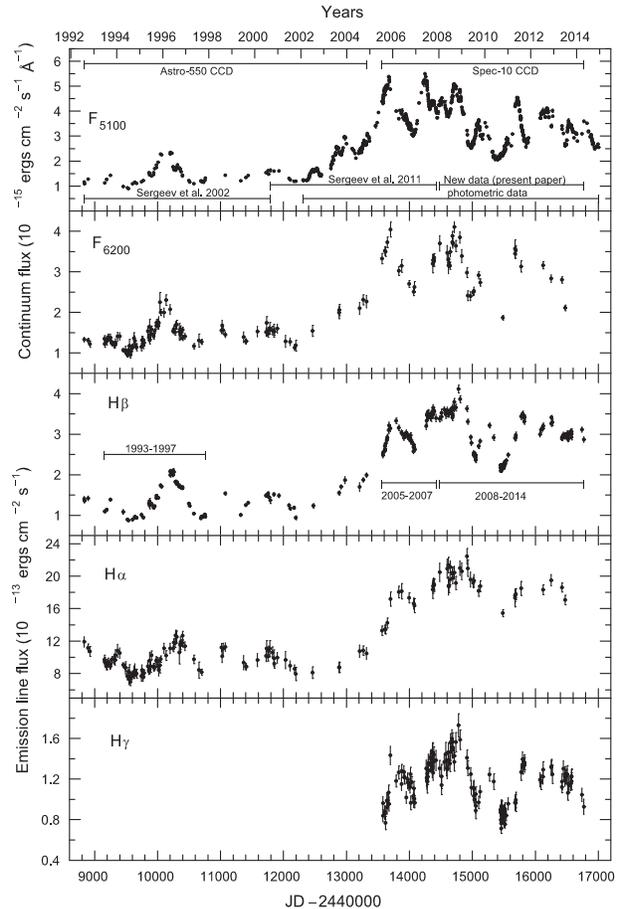}
  \caption{Light curves (top to bottom) for the continua at
$\lambda$5100\AA\ and $\lambda$6200\AA\ and for the H$\beta$, H$\alpha$, and H$\gamma$
lines. Units are $10^{-13}\,\ergs$ and $10^{-15}\,\ergsA$ for the
lines and continua, respectively.}
  \label{lc}
\end{figure}

The fluxes with their observational uncertainties are
shown in Fig.~\ref{lc} for 1992--2014.
They are presented in Tables~\ref{hbfluxes} and~\ref{hafluxes} for
2008--2014. Note, that the
$F_{5100}$ continuum flux consists of both the spectral and photometric data.
The fluxes for earlier period are given in \citet{Ser02,Ser11}.
However, the \hb-region fluxes from Table~1 of \cite{Ser11} must be
corrected for the narrow-line variability (i.e.,
divided by the factor $\varphi(t)/\varphi_0$, see eq.~\ref{eq-phi}) to be
consistent with the fluxes for 2008--2014 from
Table~\ref{hbfluxes} of the present papers.
Note that we have not separated the broad and narrow line fluxes.
Because narrow line fluxes are shown to be variable, they must slightly affect
the variability of the broad lines. However, we found this effect
to be negligible (not more than 2 per cent).

\begin{table}
  \caption{H$\beta$ region fluxes. This is a sample of the full table, which is available online.}
  \label{hbfluxes}
  \begin{tabular}{@{}lccc@{}}
  \hline
   Julian Date & $F_{5100}$ & H$\beta$ & H$\gamma$ \\
   -2,440,000 & & &  \\
  \hline
14480.199 & 42.28$\pm$0.44 &   33.77$\pm$0.70 &    13.06$\pm$0.78   \\
14483.202 & 43.98$\pm$0.44 &        ...       &       ...           \\
14497.173 & 44.46$\pm$0.35 &        ...       &       ...           \\
14503.206 & 44.25$\pm$0.49 &        ...       &       ...           \\
14508.652 & 42.89$\pm$0.72 &   34.30$\pm$0.86 &    11.39$\pm$0.93   \\
  \hline
\end{tabular}
{\\\footnotesize Units are $10^{-14}\,\ergs$ and $10^{-16}\,\ergsA$ for the
lines and continuum, respectively. The fluxes are consistent with the respective
fluxes for earlier period \citep{Ser11} when both the fluxes and uncertainties from \citet{Ser11}
are divided by the scale factor $\varphi(t)$ (see text).}
\end{table}

\begin{table}
 %\centering
 %\begin{minipage}{100mm}
  \caption{H$\alpha$ region fluxes in the two wavelength windows:
6740--7160\,\AA\ and 6777.0--7033.5\,\AA\ (observer frame).}
  \label{hafluxes}
  \begin{tabular}{@{}lccc@{}}
  \hline
   Julian Date & $F_{6200}$ & H$\alpha$        & H$\alpha$                          \\
   -2,440,000  &            & 6740--7160\,\AA  & 6777.0--7033.5\,\AA\\
  \hline
14481.250 &  37.00$\pm$1.92 &  204.9$\pm$11.3  &  173.6$\pm$9.4    \\
14600.422 &  34.66$\pm$1.86 &  209.2$\pm$11.7  &  175.7$\pm$9.7    \\
14617.480 &  33.07$\pm$1.39 &  213.6$\pm$9.6   &  179.4$\pm$7.9    \\
14620.523 &  31.54$\pm$1.93 &  188.1$\pm$11.4  &  157.8$\pm$9.4    \\
14625.422 &  32.23$\pm$1.14 &  187.1$\pm$7.4   &  157.9$\pm$6.1    \\
14643.512 &  31.50$\pm$0.88 &  195.7$\pm$6.4   &  164.9$\pm$5.1    \\
14650.441 &  34.93$\pm$1.09 &  211.1$\pm$7.6   &  177.6$\pm$6.1    \\
14681.379 &  37.30$\pm$0.96 &  203.0$\pm$6.5   &  171.6$\pm$5.2    \\
14683.445 &  38.83$\pm$1.12 &  204.3$\pm$7.1   &  173.3$\pm$5.8    \\
14687.367 &  37.00$\pm$1.14 &  197.7$\pm$7.2   &  166.9$\pm$5.8    \\
14714.395 &  41.03$\pm$1.30 &  204.3$\pm$7.8   &  172.9$\pm$6.3    \\
14738.270 &  36.39$\pm$1.38 &  191.5$\pm$8.2   &  161.6$\pm$6.7    \\
14802.266 &  38.48$\pm$1.42 &  209.9$\pm$8.8   &  177.3$\pm$7.2    \\
14833.188 &  33.89$\pm$1.62 &  205.9$\pm$10.5  &  174.4$\pm$8.7    \\
14912.594 &  29.79$\pm$1.21 &  224.5$\pm$9.7   &  189.7$\pm$8.0    \\
14927.488 &  24.13$\pm$1.20 &  209.6$\pm$10.0  &  176.8$\pm$8.3    \\
14969.477 &  24.01$\pm$1.10 &  195.8$\pm$9.2   &  165.7$\pm$7.6    \\
15010.508 &  24.90$\pm$0.76 &  192.0$\pm$6.3   &  161.3$\pm$5.1    \\
15024.430 &  25.26$\pm$0.98 &  194.3$\pm$7.7   &  162.8$\pm$6.3    \\
15098.344 &  29.15$\pm$0.90 &  181.8$\pm$6.4   &  151.2$\pm$5.1    \\
15124.367 &  27.37$\pm$0.86 &  187.7$\pm$6.5   &  156.9$\pm$5.3    \\
15482.258 &  18.69$\pm$0.53 &  154.4$\pm$4.5   &  127.5$\pm$3.6    \\
15676.539 &  34.48$\pm$1.21 &  175.9$\pm$7.1   &  145.7$\pm$5.7    \\
15677.555 &  35.72$\pm$2.17 &  173.1$\pm$11.1  &  143.7$\pm$9.0    \\
15689.504 &  35.25$\pm$1.12 &  177.9$\pm$6.7   &  147.8$\pm$5.3    \\
15772.469 &  31.32$\pm$1.37 &  185.0$\pm$8.7   &  155.0$\pm$7.1    \\
16122.402 &  31.61$\pm$0.90 &  183.3$\pm$6.2   &  154.2$\pm$4.9    \\
16252.246 &  28.34$\pm$0.93 &  194.9$\pm$7.0   &  162.3$\pm$5.7    \\
16422.424 &  28.02$\pm$0.78 &  186.1$\pm$5.6   &  154.9$\pm$4.5    \\
16472.480 &  21.11$\pm$0.70 &  170.7$\pm$5.8   &  140.6$\pm$4.7    \\
  \hline
\end{tabular}
%\end{minipage}
{\\\footnotesize Units are $10^{-14}\,\ergs$ and $10^{-16}\,\ergsA$ for the
lines and continuum, respectively. \ha\ fluxes in the last column of this table
were integrated over the same line-of-sight velocities as for \hb.}
\end{table}

\section[]{Results}
\subsection[]{Variability characteristics}

The basic variability characteristics of 3C~390.3 are summarized in the
Table~\ref{var}. We considered the four periods of observations:
1993--1997, 2005--2007, 2008--2014,
and the entire period of CCD observations of 3C~390.3 at the CrAO (1992--2014).
The motivation of the division of the data into the periods is that
the first two periods were used for the cross-correlation analysis in
\citet{Ser02,Ser11}. The 1998--2004 period has been excluded 
from the cross-correlation analysis because
poor data sampling of both the \hb\ and \ha\ lines, while
2008--2014 period represents our new (unpublished) data.
On the other hand, this division can be used to check for lag changes.
In the Table~\ref{var} the parameter $F_{var}$ is the rms fractional variability
and the parameter $R_{max}$ is simply the
ratio of the maximum to minimum flux. Both the parameters are corrected for
observational uncertainties. We also give the variability characteristics
corrected for the contribution of the starlight of
the host galaxy in the case of the continuum and the narrow-line
contamination in the case of the emission lines \citep[for our estimates of
these contributions see table~1 and text in][]{Ser02}.
We accounted for the long-term trends in the narrow-line variability
when subtracting narrow-line contamination from the broad-line fluxes
(see sect.~\ref{phot} and \ref{narvar} for more details).

The considered variability characteristics for any light curve
can only be compared to another light curve when both of them
are sampled identically.
However, there are a notable difference in the sampling of our
\ha-region and \hb-region light curves.
To avoid effect of sampling we have selected quasi-simultaneous data
points from both regions to construct the identically sampled light curves.
Their variability characteristics are given in the last five rows
of the Table~\ref{var}. As can be seen from these rows, the fractional
variability amplitude is almost the same for the continuum fluxes
at $\lambda 5100$ and $\lambda 6200$ (probably it is slightly
greater for $\lambda 5100$).  Among observed Balmer lines,
the variability amplitude is greatest in \hg, less
in \hb, and lowest in \ha.

\begin{table*}
%\centering
% \begin{minipage}{100mm}
% \scriptsize
  \caption{Variability characteristics.}
  \label{var}
  \begin{tabular}{@{}lcccccccccccc@{}}
  \hline
   & \multicolumn{3}{c}{1993--1997} & \multicolumn{3}{c}{2005--2007} & \multicolumn{3}{c}{2008--2014} &\multicolumn{3}{c}{1992--2014} \\
   Feature & Mean & $F_{var}$ & $R_{max}$   & Mean & $F_{var}$ & $R_{max}$ & Mean & $F_{var}$ & $R_{max}$ &  Mean & $F_{var}$ & $R_{max}$ \\
  \hline
$F_{5100}$              &  15.0 &  0.250 & 2.60 & 41.8  &  0.162 & 1.76 & 33.0  &  0.244 & 2.32 & 29.5 & 0.417 &  5.98 \\
${F_{5100}}^\ast$       &  10.3 &  0.365 & 4.38 & 37.1  &  0.182 & 1.90 & 28.3  &  0.285 & 2.68 & 24.8 & 0.496 & 11.50 \\
H$\beta$                &  13.9 &  0.280 & 2.35 & 31.1  &  0.107 & 1.41 & 29.7  &  0.177 & 1.92 & 25.0 & 0.350 &  4.63 \\
H$\beta^\ast$           &  12.7 &  0.308 & 2.56 & 29.6  &  0.112 & 1.43 & 28.0  &  0.188 & 2.00 & 23.5 & 0.366 &  5.18 \\
H$\gamma$               &  ...  &  ...   & ...  & 11.7  &  0.140 & 1.71 & 11.7  &  0.192 & 2.16 & ...  & ...   &  ...  \\
H$\gamma^\ast$          &  ...  &  ...   & ...  &  9.4  &  0.169 & 1.95 &  9.0  &  0.255 & 2.80 & ...  & ...   &  ...  \\
$F_{6200}$              &  14.1 &  0.183 & 2.21 &  32.3 &  0.119 & 1.59 &  31.6 &  0.174 & 2.18 & 20.0 & 0.439 &  3.96 \\
${F_{6200}}^\ast$       &  7.10 &  0.363 & 4.48 &  25.3 &  0.151 & 1.82 &  24.6 &  0.223 & 2.89 & 13.0 & 0.675 &  9.57 \\
H$\alpha$               &  94.2 &  0.138 & 1.65 & 168.1 &  0.124 & 1.33 & 193.8 &  0.068 & 1.42 & 124.1& 0.353 &  2.94 \\
H$\alpha^\ast$          &  83.8 &  0.155 & 1.76 & 155.9 &  0.131 & 1.35 & 179.8 &  0.077 & 1.47 & 112.7& 0.377 &  3.20 \\
${F_{5100}}^{\ast\ast}$ &  9.86 &  0.367 & 4.23 & 36.7  &  0.184 & 1.73 & 33.6  &  0.210 & 2.14 & 22.5 & 0.594 & 10.36 \\
H$\beta^{\ast\ast}$     &  12.1 &  0.309 & 2.54 & 29.4  &  0.123 & 1.36 & 31.2  &  0.112 & 1.59 & 21.4 & 0.453 &  4.86 \\
H$\gamma^{\ast\ast}$    &  ...  &  ...   & ...  &  9.4  &  0.227 & 2.06 & 10.5  &  0.168 & 1.69 & ...  & ...   &  ...  \\
${F_{6200}}^{\ast\ast}$ &  7.37 &  0.354 & 4.74 &  25.2 &  0.165 & 1.81 &  25.1 &  0.201 & 2.39 & 16.4 & 0.578 & 11.71 \\
H$\alpha^{\ast\ast}$    &  84.9 &  0.168 & 1.55 & 157.1 &  0.121 & 1.35 & 181.7 &  0.060 & 1.28 & 129.1& 0.358 &  2.86 \\
  \hline
\end{tabular}

\smallskip
{\footnotesize Units for the `Mean' (mean flux) columns are the same as in the
Tables~\ref{hbfluxes} and \ref{hafluxes}. Rows marked with $\ast$ or $\ast\ast$
represent data with subtracted constant contributions.
Rows marked with $\ast\ast$ (last five rows) represent identically sampled light
curves.}

%\end{minipage}
\end{table*}

\subsection{Narrow-line variability}
\label{narvar}

\begin{figure}
  \includegraphics[width=55mm]{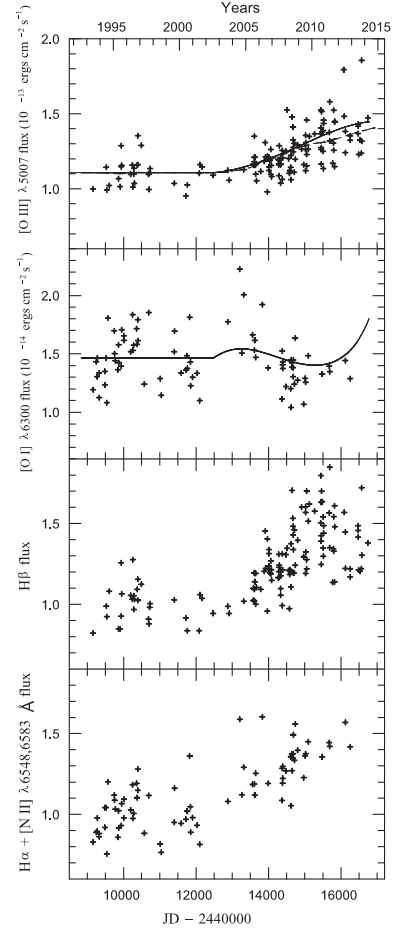}
  \caption{Shown by the crosshair symbols are the light curves
of the following narrow lines (top to bottom):
\oiii\,$\lambda5007$, \oi\,$\lambda6300$, \hb, and
\ha\,+\,\nii\,$\lambda\lambda 6548,6583$.
The narrow-line fluxes were computed
from the spectra obtained under photometric
conditions by using the spectra of the comparison star.
Flux units in the last two panels are arbitrary.
The suspected variations of the \oiii\ and \oi\ lines according to
the changes in the scale factor $\varphi$ are shown by the solid lines,
and the predicted variations in the \oiii\ line in terms of the model that is
considered as an example in \citet{Ser11} are shown by the dashed line.}
  \label{oiii}
\end{figure}

The discrepancy between our spectral and photometric data sets (sect.~\ref{phot})
can be attributed to the long-term variability of the
\oiii\,$\lambda\,5007$\,\AA\ forbidden line.
If so, the light curve of the \oiii\ line must be inversely proportional
to the scale factor $\varphi(t)$ that was determined in sect.~\ref{phot}.
To check the narrow-line variability in 3C~390.3 the
narrow-line fluxes were computed for photometric nights
using the absolute flux calibration by the comparison star.
To reduce some contribution from the
far red wing of the \hb\ line to the \oiii\ line
and from the far blue wing of the \ha\ line to the \oi\,$\lambda$6300 line
we selected a slightly narrower
integration zones for the narrow lines as compared to \cite{Ser02,Ser11}
with closely adjacent pseudo-continuum windows
(both red-side and blue-side windows for each line).
The light curves of the \oiii\ and \oi\
lines are shown in Fig.~\ref{oiii}. Also given in Fig.~\ref{oiii} are inverse
scale factors $\varphi(t)^{-1}$  for both lines.

We determined the probability that
the observed narrow-line fluxes are proportional to $\varphi(t)^{-1}$
(alternative
statistical hypothesis is that the narrow-lines do not vary in  flux).
We have found, from the $\chi^2$-criterion, that the constant flux hypothesis must be rejected
at $0.94$ and $1-1.4\times 10^{-9}$ for the \oi\ and \oiii\ lines,
respectively.
So, the confidence level for the correlation between narrow-line fluxes
and $\varphi(t)^{-1}$ is almost 100 per cent
for the \oiii\ line, but it is not significant enough for the \oi\ line.

In \citet{Ser11} it was shown that a shell model of the NLR
with inner and outer radii of 20 and 200 light-years, respectively,
and the emissivity per unit volume
$\varepsilon \propto r^{-3.2}$ gives a good fit to
the changes in the \oiii\ line flux (see their Fig.~3).
Note that this model has been considered as an example, not a fit.
It is remarkable that it gives a good prediction
for the \oiii\ variability in 2008--2014 (top panel of Fig.~\ref{oiii}).
After several attempts we have found parameters of a shell models
that provide an ideal fit to $\varphi(t)^{-1}$
with rms agreement to be less than one per cent, that is:
inner and outer radii of 20 and 150 light-years, respectively,
and the emissivity per unit volume
$\varepsilon \propto r^{-3.5}$. We conclude that there are a lot of NLR models
that is able to reproduce the observed \oiii\ variability.

The observed variability of the \oiii\,$\lambda\,5007$ forbidden line
strongly suggests that the narrow Balmer lines must be variable in flux as well.
However, to measure fluxes of the narrow Balmer lines,
they must be separated from the respective broad lines.
Estimates of the narrow Balmer line fluxes were obtained via finding
a scale factor for each spectrum to achieve a maximum smoothness of the
residuals between a given spectrum and template spectrum
\citep[see calibration method by][independently developed in the CrAO]{Gro92}.
In other words,
there must be no narrow-line residuals seen
over spectral region where these narrow lines reside
when the template spectrum is subtracted from a given spectrum.
As a criterion for smoothness we have selected the rms agreement
between residuals for a given spectrum and smoothed version of these residuals
(only for a spectral region where a given narrow line resides).
To smooth residuals we have used a convolution with $\Pi$-shaped function
with 24\,\AA\ full width
(i.e., the function value is a positive number in the $\pm 12$\,\AA\
interval, and it is zero outside this interval).
The results for the \hb\ narrow line and for the
\ha + \nii\ complex of narrow lines are shown in Fig.~\ref{oiii}.
It is clearly seen that above lines vary in flux and that these
variations are similar to the \oiii\ line variations.

We have compared the variability amplitudes
of the \hb, \ha + \nii, and \oiii\ narrow lines.
We considered long-term flux changes only, since
a large short-term scattering of the observed narrow-line fluxes
is the observational uncertainties.
We found an rms amplitude to be about $F_{var}=0.14$,
and max--to--min variations to be a factor of $R_{max}=1.6$
for both the \hb\ and \ha + \nii\ lines.
The \oiii\ line shows smaller variability amplitude of
$F_{var}=0.083$ and $R_{max}=1.36$.

We have experimented with various criteria for smoothness of the residuals
(various widths of the $\Pi$-shaped kernel or low-order polynomial functions
as a smoothed version of the residuals) and we found similar variability
patterns of the \hb\ and \ha + \nii\ narrow lines with the same or slightly higher
(up to a factor of 1.3) variability amplitudes.

\begin{figure*}
  \includegraphics[width=140mm]{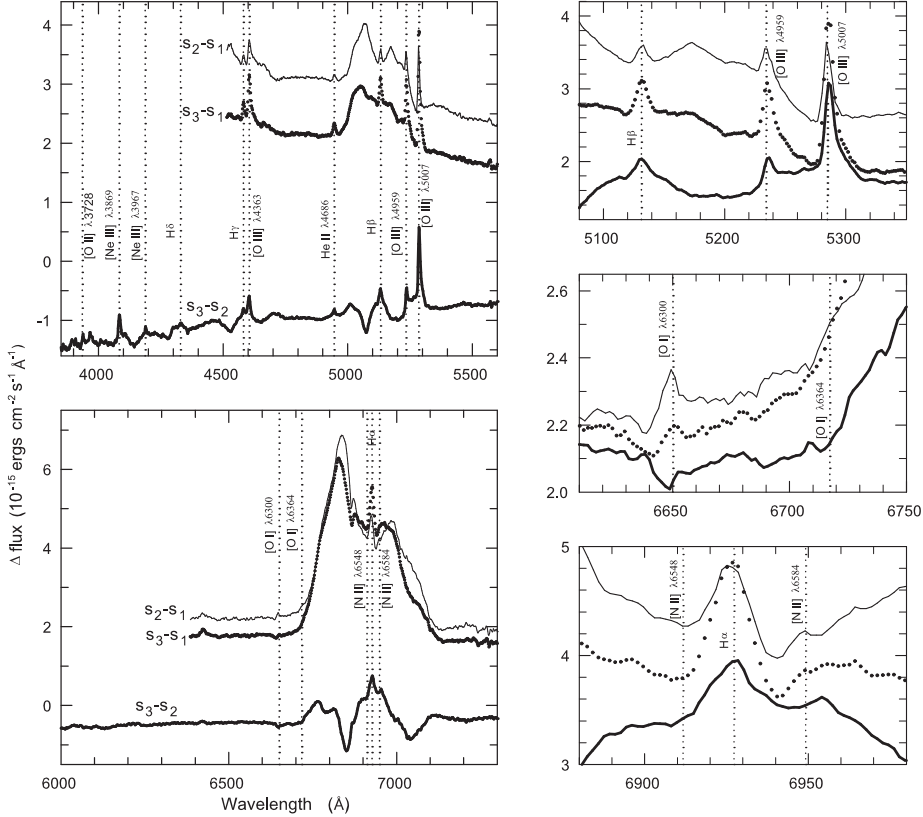}
  \caption{Residuals between mean star-calibrated spectra
  ($S_1$, $S_2$, and $S_3$) for 1993--1997,
  2005--2007, and 2008--2014, respectively.
  The $S_2$-$S_1$, $S_3$-$S_1$, and $S_3$-$S_2$ flux differences are shown
  by the thin solid line, filled circles, and thick solid line, respectively.
  Right panels show an enlarged portion of wavelengths near \hb\ and
  \oiii$\lambda5007$ (top), \oi$\lambda6300,6364$ (middle), and \ha+\nii\
  (bottom).  The residuals in the right panels are artificially
  shifted in Y-axis for better view.
  Dotted vertical lines indicate position of the
  strong narrow lines.   Narrow-line features in the residuals
  are under-subtracted narrow lines.}
  \label{res}
\end{figure*}

Finally, in Fig.~\ref{res} we show the differences in fluxes between mean
star-calibrated spectra $S_1$, $S_2$, and $S_3$
for the three periods: 1993--1997, 2005--2007, and 2008--2014, respectively
($S_2$-$S_1$, $S_3$-$S_1$, $S_3$-$S_2$ flux differences are shown).
The residuals in Fig.~\ref{res} clearly demonstrate many narrow-line features
associated with the under-subtracted narrow lines.
It proves that the narrow lines are indeed variable.
It is clearly seen that most of the narrow lines in Fig.~\ref{res}
are brighter in 2005--2007 than in 1993--1997,
and brighter in 2008--2014 than in 2005--2007.

\subsection[]{Time-series analysis}

\begin{figure}
  \includegraphics[width=60mm]{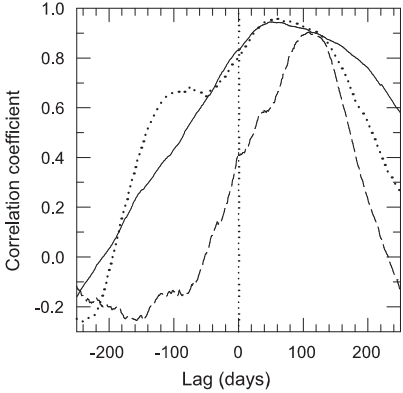}
  \caption{Continuum--\hb\ cross-correlation functions for 1993--1997 (solid line),
2005--2007 (dashed line), and 2008--2014 (dotted line). Vertical dotted line
indicates zero lag.}
  \label{ccf}
\end{figure}

The time delays between various light curves were determined
as in \citet{Ser11}, i.e., using the interpolation cross-correlation
function \citep[ICCF, e.g.,][]{gas86, wh94}.
The results of the cross-correlation analysis for 2008--2014
are shown in Table~\ref{cor}.
Also given in Table~\ref{cor} are the cross-correlation results from the previous
Crimean studies: for 1993--1997 \citep{Ser02} and for 2005--2007 \citep{Ser11}.
The continuum--\hb\ CCFs for the considered periods are shown in Fig.~\ref{ccf}.

\begin{table*}
\begin{minipage}{42pc}
  \scriptsize
  \caption{Cross-correlation results.}
  \label{cor}
  \begin{tabular}{@{}lccccccccccccc@{}}
  \hline
 Series  & \multicolumn{4}{c}{1993--1997} & \multicolumn{4}{c}{2005--2007} & \multicolumn{4}{c}{2008--2014} & Mean\\
 & $\tau_{peak}$ & $\tau_{cent}$ & $r_{max}$ & $P$
 & $\tau_{peak}$ & $\tau_{cent}$ & $r_{max}$ & $P$
 & $\tau_{peak}$ & $\tau_{cent}$ & $r_{max}$ & $P$ & $\tau_{cent}$\\
\hline
 $F_{5100}$ -- H$\beta$   & $50^{+21}_{-2}$   & $82^{+8}_{-8}$    & 0.945 & 0                 & $108^{+6}_{-11}$  & $110^{+3}_{-9}$   & 0.903 & 0     & $56^{+14}_{-1}$  & $74^{+5}_{-3}$    & 0.960 & 0       & $88.6\pm8.4$ \\
 $F_{5100}$ -- H$\alpha$  & $196^{+16}_{-79}$ & $162^{+31}_{-12}$ & 0.848 & 0                 & $183^{+23}_{-40}$ & $179^{+13}_{-20}$ & 0.936 & 0     & $117^{+65}_{-5}$ & $144^{+25}_{-26}$ & 0.839 & 0.002   & $161\pm15$   \\
 $F_{5100}$ -- H$\gamma$  & ...               & ...               & ...   &...                & $136^{+5}_{-12}$  & $134^{+7}_{-11}$  & 0.806 & 0     & $60^{+59}_{-1}$  & $91^{+8}_{-7}$    & 0.912 & 0       & $113\pm14$   \\
 H$\beta$   -- H$\alpha$  & $56^{+55}_{-3}$   & $80^{+22}_{-13}$  & 0.922 & $4\times 10^{-4}$ & $49^{+66}_{-12}$  & $82^{+26}_{-29}$  & 0.972 & 0.005 & $51^{+62}_{-13}$ & $72^{+49}_{-20}$  & 0.843 & 0.003   & $78\pm15$    \\
 H$\beta$   -- H$\gamma$  &  ...              & ...               & ...   &...                & $37^{+8}_{-17}$   & $32^{+25}_{-12}$  & 0.925 & 0.011 & $19^{+16}_{-14}$ & $16^{+13}_{-5}$   & 0.947 & 0.010   & $24\pm12$    \\
 $F_{5100}$ -- $F_{6200}$ & $-8^{+11}_{-10}$  & $-7^{+11}_{-12}$  & 0.947 & 0.74              & $27^{+12}_{-26}$  & $20^{+10}_{-14}$  & 0.985 & 0.075 & $1^{+3}_{-5}$    & $-5^{+5}_{-4}$    & 0.970 & 0.82    & $2.6\pm7.4$  \\
  \hline
\end{tabular}
NOTE: $P$ is a probability that $\tau_{cent}$ is less than zero.
H$_\alpha$ line fluxes are integrated over the same line-of-sight velocities as for H$_\beta$.
\end{minipage}
\end{table*}

The narrow-line variability leads
to the appearance of fictive long-term trends in the
derived light curves.
The cross-correlation results as well as other results
of the present paper are given
with accounting for this trend for the
 \hb-region light curves (\hb, \hg, and $F_{5100}$),
including 2005--2007 period of observations.
See sect.~\ref{phot} and \ref{narvar} for more details.

\begin{figure*}
  \includegraphics[width=140mm]{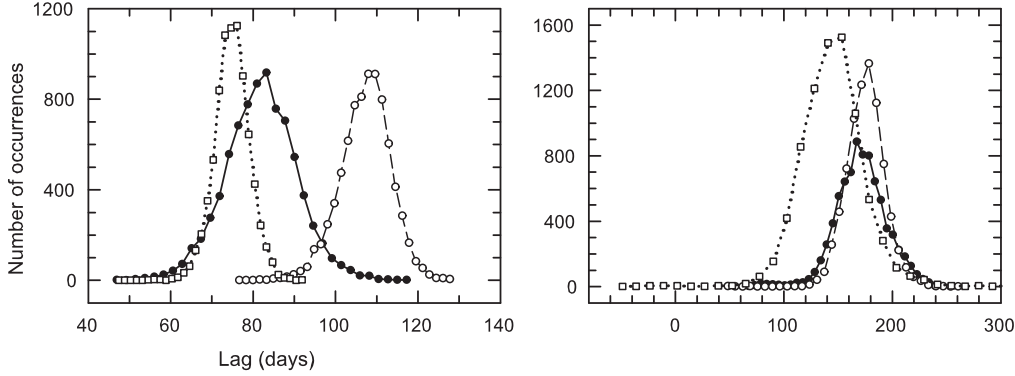}
  \caption{Probability distributions for a lag ($\tau_{cent}$) between
  continuum  and both \hb\ (left panel) and \ha\ (right panel) lines
  for 1993--1997 (filled circles connected by a solid line), 2005--2007
 (open circles connected by a dashed line), and 2008--2014
 (open squares connected by a dotted line). The distributions are derived
  from the FR/RSS technique.}
  \label{hst}
\end{figure*}

The lags were measured
from both the location of the maximum value of the ICCF correlation coefficient
($r_{max}$) and from the ICCF centroid based on all points with $r \geq 0.8r_{max}$
(designated as $\tau_{peak}$ and $\tau_{cent}$, respectively).
The lag uncertainties were computed using the model-independent Monte Carlo flux
randomization/random subset selection (FR/RSS) technique described by
\citet{Pet98a}. From this technique we have obtained the lag probability
distributions. Fig.~\ref{hst} shows probability distributions for $\tau_{cent}$
for both \hb\ and \ha\ lines (respective lag measurements
are in the first two rows of Table~\ref{cor}).
We found that the lag for the \hb\ line in 2005--2008 is different from
that in 1993--1997 and in 2008--2014 with a probability of 0.989 and 0.9997,
respectively. We have conservatively attributed above difference to
some additional lag uncertainty (see below). No significant difference
in the \ha\ lag among considered periods has been found.

Also given in Table~\ref{cor} are unweighted mean lags for three periods
of observations. The uncertainties in the mean lag were computed
via maximum likelihood method
in assumption that there is an
internal lag scattering in addition to the scattering derived from the
FR/RSS technique for the individual observational periods.
A reason to introduce this internal uncertainty is
clearly seen from the Fig.~\ref{hst}: the computed probability distribution
for the \hb\ lag is too different between 2005--2007 and 2008--2014,
so an additional uncertainty is required. This uncertainty can be attributed
to a lot of unaccounted factors, e.g., real lag changes, underestimated
observational uncertainties, etc.
The likelihood method is widely used
for constructing estimators of unknown parameters in statistics.
Maximum-likelihood estimation was vastly popularized by Ronald Fisher
\citep[e.g.][]{Fish1912}.
In our implementation of this method, an additional lag scattering $\sigma$
is unknown parameter to be find by achieving a maximum of the function:
\[ L(\sigma) =   \prod_{i=1}^{3} \rho_i(\tau=\tau_0,\sigma)~,\]
where the index $i$ represents one of the three observing periods,
and $\rho_i(\tau,\sigma)$ is a convolution of the two probability distributions
of $\tau$:
one is obtained from the FR/RSS technique for a given period (see Fig.~\ref{hst}),
while another is a normal distribution with unknown standard deviation $\sigma$
and with the expectation equals to the mean lag $\tau_0$
from the last column of Table~\ref{cor}.
Such a method provides larger lag uncertainties than the FR/RSS technique
does. E.g., the uncertainties in the mean lag can even be greater
than the uncertainties for individual periods of observations when
there is a large scattering in individual lag values.
The additional uncertainty in the lag between the continuum and \hb\
emission was found to be $\sigma=13$\,d.

\subsection[]{Central black hole mass}
\label{bhmass}

The reverberation mass of the central black hole is given by
$$
M_{BH}=\frac{fc\tau\sigma^2_{line}}{G}\,,
$$
where $G$ is the gravitational constant, $c$ is the light velocity,
$\tau$ is the emission-line time delay (a proxy for the BLR size),
$\sigma_{line}$ is the width of this line (a proxy for the velocity dispersion
of BLR gas), and $f$ is the scaling factor of order
unity that depends on the BLR geometry, kinematics, and orientation.
To obtain the $M_{BH}$ directly in
the Solar mass units, the above equation can be written as
\begin{equation}
M_{BH}=\frac{k\tau\sigma^2_{line}}{1.14\times10^{-10}}\,,
\label{mass}
\end{equation}
where $k\equiv2f$, $\sigma_{line}$ is in light-velocity units, $\tau$ is
in days, and $1.14\times10^{-10}$ is gravitational radius of Sun in light-days.
Here we adopt a definition for the gravitational radius $r_g$
according to \citet{Opp39} \citep[the same definition was used in][]{Ser02,Ser11}, i.e., $r_g = 2\,G\,M / c^2$.

While it is difficult to determine $f$ for an individual
source because dynamical modeling is required,
its average value has been determined by comparing
the available reverberation measurements with the measurements
from the well known $M_{BH}-\sigma_*$ relationship for AGNs.
The most recent determination
of the scale factor for reverberation-mapped AGNs is
$f = 4.31\pm1.05$ \citep{Gri13} and it is
consistent with earlier results \citep[e.g.,][]{Onk04,Gra11}.
As in \citet{Ser11} we adopt $f$ to be $5.5$ according to
\citet{Onk04}.

Recent results of observations obtained with high sample rate show the
diversity and probable complexity of the BLR velocity field \citep{den09,Den10,Gri12}
with indications of outflowing component at least for NGC~3227 nucleus.
Since the reverberation method is based on the virial assumption,
it does call into question the reliability of this method
for the determination of black hole masses.

\begin{table*}
  \caption{\ha\ and \hb\ profile width.}
  \label{width}
  \begin{tabular}{@{}lccccccccc@{}}
  \hline
           & \multicolumn{3}{c}{1993--1997} & \multicolumn{3}{c}{2005--2007} & \multicolumn{3}{c}{2008--2014}\\
 Component & FWHM           & $\sigma_{line}$ & Centroid  & FWHM    & $\sigma_{line}$ & Centroid & FWHM     & $\sigma_{line}$ & Centroid \\
           & (km\,s$^{-1}$)  & (km\,s$^{-1}$)   &   (\AA) & (km\,s$^{-1}$)  & (km\,s$^{-1}$)   &   (\AA) & (km\,s$^{-1}$)  & (km\,s$^{-1}$)   &   (\AA)  \\

  \hline
\ha\ mean      & 11380 & 4310 &  6932.0  & 10600 & 4180 & 6910.7 & 10490 & 4150 & 6909.5\\
\hb\ mean      & 12690 & 4860 &  5143.8  & 12520 & 4280 & 5132.2 & 13740 & 4700 & 5132.3\\
\ha\ rms       & 11570 & 4270 &  6935.8  &  8830 & 4010 & 6901.4 & 9480  & 4150 & 6922.5\\
\hb\ rms       & 12010 & 4520 &  5142.2  & 12330 & 4530 & 5121.7 & 9460  & 4620 & 5123.6\\
  \hline
\end{tabular}
\end{table*}

The line width is usually measured as FWHM or line dispersion $\sigma_{line}$
(i.e. the second moment of the line profile).
It has been argued that $\sigma_{line}$ is a less biased measure of
the velocity dispersion of BLR gas than FWHM \citep[e.g.,][]{Ser99,Col06}.
Similarly, it was argued that the $\tau_{cent}$ above the correlation
coefficients of $0.8r_{max}$ is the most representative lag estimate for the
BLR radius \citep{pet04}. It seems, from our experience, that the $\tau_{cent}$ distribution
is more normal with slightly less deviations than that of $\tau_{peak}$.
Estimates of $\tau_{cent}$ above lower correlation levels
seem to be not applicable because far CCF wings are poorly determined
since a limitation on time shift.
Note that a significant difference between the
$\tau_{peak}$ and $\tau_{cent}$ is only pronounced if the BLR is strongly
extended spatially, so the transfer function has a long tail.

In Table~\ref{width} we give the \ha\ and \hb\ broad-line widths
and centroids
(narrow emission-lines are removed from the broad profiles)
for the three periods of observations: 1993--1997, 2005--2007, and 2008--2014.
They were computed from both the mean and rms profiles and
for the same time intervals as that of the time
lag computations. The uncertainties in $\sigma_{line}$ were computed
as described in \citet{Ser11} and they were found to be about 5 per cent
for all observing periods, while the uncertainties in the centroids
were found to be 1.1\,\AA\ and 2.5\,\AA, for the \hb\ and \ha\ lines
respectively.
%We found that the accounting for the narrow-line variability
%adds too little to the previously published line widths
%\citep[before 2008, see][]{Ser02,Ser11} and these widths were left
%unchanged.

As can be seen from Table~\ref{width}, the $\sigma_{line}$ value
for both lines is remarkably constant among all periods.
It seems that the widths of the Balmer lines in 3C~390.3
(as well as their lags, see Table~\ref{cor}) depend little,
if any, on the continuum flux that
was much brighter in 2005--2014 than in 1993--1997.

Also seen from Table~\ref{width} is that the \ha\ and \hb\ broad-line
centroids have shifted to shorter wavelengths between 1993--1997 and 2005--2014
because systematic brightening of the blue bump of the line profiles.
There were no changes in the centroids between 2005--2007 and 2008--2014
for both the \ha\ and \hb\ lines except for the rms profile of the \ha\ line
that is shifted to longer wavelengths.

To compute the average mass for the three observational periods
we adopt the $\sigma_{line}$ value for the rms profiles
to be a mean value among three periods
(note, that these three values are almost the same).
The mean $\sigma_{line}$ value for the three periods from Table~\ref{width}
is equal to 4557 and 4143\,km\,s$^{-1}$, while
the mean $\tau_{cent}$ value
is equal to $88.6\pm8.4$ and $161\pm15$\,d
(see last column of Table~\ref{cor}) for the \hb\ and \ha\ lines, respectively.
The lag values need to be corrected
for time dilation by dividing by $1+z$ to put them into the rest frame.
After correction for time dilation (assuming $z=0.05553$)
the corresponding estimates of the black hole mass in
3C~390.3 are $(1.87\pm0.26)\times10^9\,M_\odot$ and
$(2.81\pm0.38)\times10^9\,M_\odot$ under $f=5.5$.
Here we have conservatively adopted the uncertainties in the mean width
of both lines to be equal to the uncertainties for the individual periods,
i.e, 5 per cent.
Note, that the reverberation mass uncertainties originate not only from the observational
uncertainties and data sampling, but from the uncertainties in the scale
factor $f$, and from the validity of the virial assumption for a given emission-line
in a given AGN.

As was shown by \cite{Ser99} the coefficient $k$ in eq.~\ref{mass} is approximately
equal to $4/\sin^2i$ for the Keplerian disc model, where $i$
is the inclination angle ($i=0\degr$ for pole-on configuration).
Assuming the BLR in the 3C~390.3 nucleus
to be a Keplerian disc with $i=27.6\degr$ \citep[see][]{Ser02}, the expected
value of the scale factor is $f\approx 9.3$. If so, our reverberation
mass must be enlarged by a factor of $9.3/5.5 = 1.7$.
Indeed the model-depended mass estimate in sect.~\ref{kepler} gives a
larger black hole mass in 3C~390.3.

\citet{Wan99} and \citet{Die12} claimed the central black hole
in 3C~390.3 to be $4\times10^8\,M_\odot$ and $8.6\times10^8\,M_\odot$, respectively.
This is several times less than our mass estimate.
Except for a difference in both the adopted scale factor and
in the adopted method to measure the line width, the principal
reason for above disagreement is a disagreement in the measured \hb\ lag.
This lag was found to be 24 days and 44 days in \citet{Wan99} and in \citet{Die12},
respectively.
\citet{Ser02} have argued that this disagreement in the lag must be attributed to
the broader ACF width for our time series \citep[see also][to learn more about
how the ACF width affects lag measurements]{Net90}.
The ACF width is expected to be systematically greater for a longer
observing campaign because long-term program shows variability timescales
not sampled in the short-term campaign.
Our monitoring campaign is definitely longer than that of
\citet{Wan99} and \citet{Die12} (about one year and several months, respectively).
In contrast to the short-term campaigns above, the multi-year campaigns
give the \hb\ lag in 3C~390.3 to be comparable to our lag estimate.
Thus \citet{Sha10} claimed a lag of 95 days for the \hb\ line
and 120 days for the \ha\ lines,
while \citet{Afa15} found this lag to be 60--79 and 138--186 days, respectively.
It is clear that the mass of the black hole determined from above lags
(the authors did not calculate the mass)
should be in agreement with our mass estimate.

The central black hole mass in 3C~390.3 was found to be
$(5\pm 1)\times10^8\,M_\odot$ from the stellar velocity dispersion
\citep{Lew06}, again less than our mass estimate.
This disagreement can be attributed to the intrinsic scattering
in the $\sigma_* - M_{BH}$ relationship as well as to uncertainties
in the scale factor $f$. These uncertainties are related to the
unknown geometry/kinematics and to a possible non-virial gas motions
in the BLR.

Since our new $\tau_{cent}$ and $\sigma_{line}$ measurements
for 2008--2014 are almost the same as in \citet{Ser11},
it is not surprising that the mass estimates are almost the same as well,
and that the 3C~390.3 nucleus is indeed an `outsider'
in the fundamental relationship between
optical luminosity and black hole mass.
Indeed, we found, from all observing periods, that the Eddington
ratio is $E_{bol}/E_{Edd} = 0.0037$. Here we adopt
$L_{bol} = 1.3\times 10^{38} M_{BH}$ and
$L_{Edd} = 9 \times \lambda\,L_{5100}$, where $L_{bol}$ is the
bolometric luminosity, $L_{Edd}$ is the Eddington luminosity, and
$L_{5100}$ is the rest frame absolute luminosity of the 3C~390.3 nucleus
at $\lambda = 5100$\AA\ (in units of ergs\,\AA$^{-1}$).
The bolometric luminosity has been corrected for Galactic reddening
using $A_{V} = 0.22$ according to the reddening map from \citet{Sch98}
and it was found to be $L_{bol} = (1.01\pm 0.25) \times 10^{44}$\,ergs.

\subsection{Model-dependent black hole mass}
\label{kepler}

In \citet{Ser11} the black hole mass in
3C~390.3 has been derived in terms of the relativistic Keplerian disc
model of BLR \citep[e.g.,][]{ch89b, er94}. The disc parameters
(the inner/outer radii, inclination angle, and emissivity law per unit surface)
were taken from \citet{Ser02}.
The free parameters are: the black hole mass $M_{BH}$,
the power-law index $a$, and a scale factor $c$ \citep[see][for more details]{Ser11}.
These parameters were obtained by fitting of modeled light curves of the \hb\ and
\ha\ emission-lines to the observed light curves.
It was assumed that the source of the continuum that drives the line is
slightly above a disc centre (so its height is much less than the
disc internal radius)
and both the line and the optical continuum fluxes are power-law functions of the
driving continuum fluxes, so the relationship between
the line fluxes and optical continuum fluxes is a power-law as well:
$F_{line} = c F_{cont}^a$.  We select optical continuum
light curve to be $F_{5100}$ (see Fig.~\ref{lc}, top panel).
In Table~\ref{modmass} we give the fit results
for our new data set in 2008--2014 as well as for earlier periods and for
the entire period of 1992--2014.
Note that the masses in Table~\ref{modmass} should be divided
by $1+z$ because observed light curves are not corrected for time dilation.

\begin{table}
  \caption{Determination of the black hole mass in terms of the Keplerian disc model}
  \label{modmass}
  \begin{tabular}{@{}lcccc@{}}
  \hline
  Years  &  $M_{BH}$ & $a$  &  $ \frac{\chi^2_{fit}}{dof}$ &  $P(\chi^2>\chi^2_{fit})$ \\
  \hline
  \multicolumn{5}{c}{\hb\ line fit}\\
  1993--1997     & $2.2\times 10^9$    & 0.80   &  4.47    & $2\times 10^{-22}$ \\
  2005--2007     & $4.1\times 10^9$    & 1.12   &  1.85    & $1.4\times 10^{-4}$ \\
  2008--2014     & $1.9\times 10^9$    & 0.68   &  4.29    & $0$\\
  1992--2014     & $3.2\times 10^9$    & 0.71   & 14.2     & $0$\\
  1992--2014$^a$ & $2.5\times 10^9$    & 0.77   &  4.77    & $0$\\
  \multicolumn{5}{c}{\ha\ line fit}\\
  1993--1997     & $4.3\times 10^9$    & 0.44   &  1.48    & $0.0072$ \\
  2005--2007     & $6.1\times 10^9$    & 1.04   &  0.49    & $0.923$ \\
  2008--2014     & $4.2\times 10^9$    & 0.43   &  1.76    & $0.0086$ \\
  1992--2014     & $9.4\times 10^9$    & 0.54   &  4.08    & $0$\\
  1992--2014$^a$ & $4.1\times 10^9$    & 0.54   &  1.40    & $0.0016$\\
  \hline
\end{tabular}
\\
{\footnotesize $^a$ With a long timescale trend removed, see text.}
\end{table}

From the inspection of the Table~\ref{modmass}
the same conclusions as from the respective table in \citet{Ser11} can
be made, namely:

\begin{enumerate}
\item The fit quality in term of the $\chi^2$ is too poor in many cases,
especially for the entire period of observations, so the probability computed
from the high tail of $\chi^2$ distribution (see the last column of
Table~\ref{modmass}) is too low. The lower $\chi^2$
values for the \ha\ line is mostly due to larger uncertainties of its light
curve.
\item There is a discrepancy in determination of
the $M_{BH}$ between  \ha\ and \hb\ lines as well as between 1993--1997 and
2005--2007.
\item The \ha\ line shows lower values of the power-law index $a$ as
compared to the \hb\ line, except for 2005--2007 when the responses of both
lines to the optical continuum variations seem to be almost linear.
\end{enumerate}

As was shown by \citet{Ser11}, the line responses to the
variations of the continuum fluxes are non-linear (except for 2005--2007),
especially for the \ha\ line.

\begin{figure}
  \includegraphics[width=80mm]{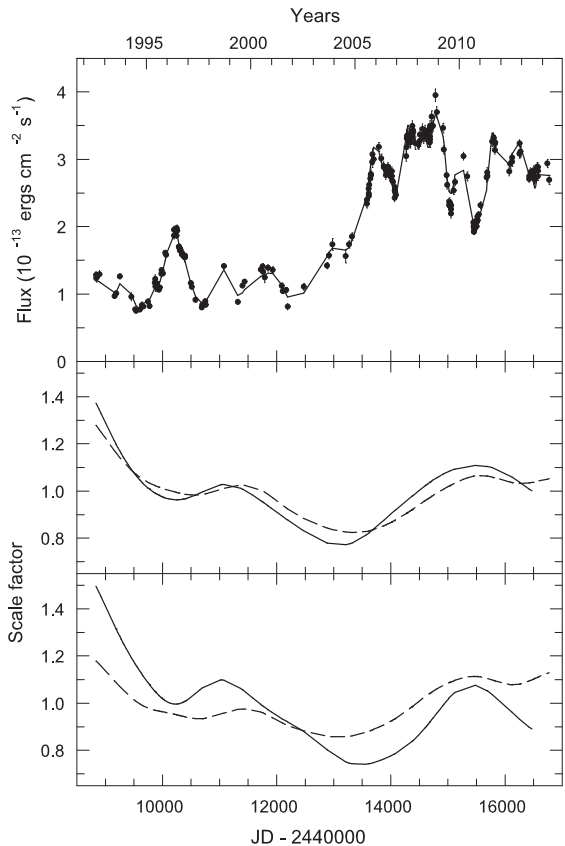}
  \caption{Observed \hb\ light curve is shown by filled circles in the
  top panel. The fit to this curve in terms of
  the Keplerian disc model (see text) is shown by solid line.
  The middle and bottom panels show long-term changes in the
  scale factor $c$ in the relationship $F_{line} = c\,F_{cont}^a$
  between line and continuum fluxes
  for the \hb\ and \ha\ lines (dashed and solid lines, respectively).
  The middle panel shows this factor
  as obtained from the Keplerian disc model, while the bottom panel shows
  this factor as obtained from the model-independent calculations (see text).}
  \label{fitfig}
\end{figure}

To improve the fit quality for the entire period, we assumed that
there is a long-term trend in the line--continuum relationship.
The long-term changes in this relationship were found, for example, by \citet{Mal97}.
To account for such changes, \citet{Den10} have proposed to use low-order
polynomials to remove long-term trends from the light curves before
applying cross-correlation analysis.
We supposed that the poor fit quality for the entire period
is related to the slow changes
in the scale factor $c$, so that this factor is a function of time and
$F_{line} = c(t) F_{cont}^a$. To remove this effect the fit procedure has been
modified as follows: (1) the modeled light curve is divided by the observed light
curve; (2) the division result is smoothed by convolution
with the Gaussian function with $\sigma=1$\,year;
(3) the modeled light curve is divided by the smoothed curve to represent the
observed light curve.
The large improvement of the modified fit for the entire period is clearly
seen from the Table~\ref{modmass}: the $\chi^2$ values per d.o.f decrease
from 14.2 to 4.77 for the \hb\ line and from 4.08 to 1.40 for the \ha\ line.
In Fig.~\ref{fitfig} we show both the modeled and observed light curves of the
\hb\ line and the normalized scale factor $c(t)$ for both \ha\ and \hb\ lines.
There is a close similarity between both scale factors.
They have reached a minimum in 2004, i.e., approximately at the beginning
of the systematic continuum brightening.

Long-term variations
of the scale factor can be interpreted, for example, via changes in the
height of the driving continuum source above the disc centre.
In terms of the considered disc model, the incident flux of the driving continuum
has to be proportional
to the height of its source for distances much exceeding this height.

Even after the modified fit is applied, the $\chi^2$ values for both lines
are significantly greater than their expected values. There are a lot of reasons
for it, e.g.: (1) the considered disc model is too oversimplified or it is not correct;
(2) the long-term
changes in the line--continuum relationship can be more complex than it is
proposed and this relationship is not necessary to be a power-law;
(3) the observational uncertainties can be underestimated; and (4) there are effect
of gaps in the continuum light curve.
In particular we found that the reasonable filling the largest gaps
in the continuum light curve (made by eye) reduces the $\chi^2$ per d.o.f. value for the \hb\
line from 4.77 to 3.45.

We have verified whether similar long-term variations in the scale factor
can be obtained by a model-independent way and how much these variations
affect the cross-correlation results for the entire observational period.
We used a modified cross-correlation function in which the continuum
light curve is adjusted to the line light curve
by the same method as in our modified fitting procedure except
for the modeled light curve of a line is now the continuum light curve
shifted by a given time lag. So, for each lag value we must perform the
adjustment before compute the correlation coefficient.
We computed the cross-correlation function between $F_{line}$ and
$F_{cont}^a$ for a set of power-law indices $a$
in order to find a largest $r_{max}$ value. We found that the power-law indices
of 0.67 and 0.66 provide largest $r_{max}$ values for the \hb\ and \ha\ lines,
respectively. For above indices, the $\tau_{cent}$ value was found
to be $80_{-4}^{+4}$\,d for \hb\ and $164_{-16}^{+12}$\,d for \ha\ with
$r_{max}$ values equal to 0.984 and 0.981, respectively. The cross-correlation
function with no correction for the variations in the scale factors but with
the power-law version of the continuum light curve gives the $\tau_{cent}$
value of 95\,d with  $r_{max}=0.954$ for the \hb\ line and of 190\,d
with  $r_{max}=0.939$ for the \ha\ line. For linear relationship between
line and continuum fluxes, the cross correlation results are as follows:
$\tau_{cent}=101$\,d and $r_{max}=0.942$ for the \hb\ line and
$\tau_{cent}=194$\,d and $r_{max}=0.938$ for the \ha\ line.
Because CCFs for both lines for the entire time interval is too flat-top
the centroid at $0.8\,r_{max}$ can not be computed. Instead, to integrate over
lag values, we used the lag intervals comparable to that for individual
periods (about 160 and 320 days for the \hb\ and \ha, respectively). These
intervals correspond to about $0.96\,r_{max}$.
Note a decrease in the derived lags and a
significant increase in the correlation for both lines after accounting for
variations in the scale factors.

Our Keplerian disc model gives the correlation coefficients
between observed and modeled light curves of  $0.990$ and $0.988$ for
the \hb\ and \ha\ lines, larger than
the correlation coefficients of the modified CCF above.

The derived scale factors $c(t)$ for both lines are shown in the bottom panel
of Fig.~\ref{fitfig}. There is a similarity between scale factors
from the Keplerian disc model and from the model-independent calculations.

\subsection{Balmer decrement}

\begin{figure}
  \includegraphics[width=60mm]{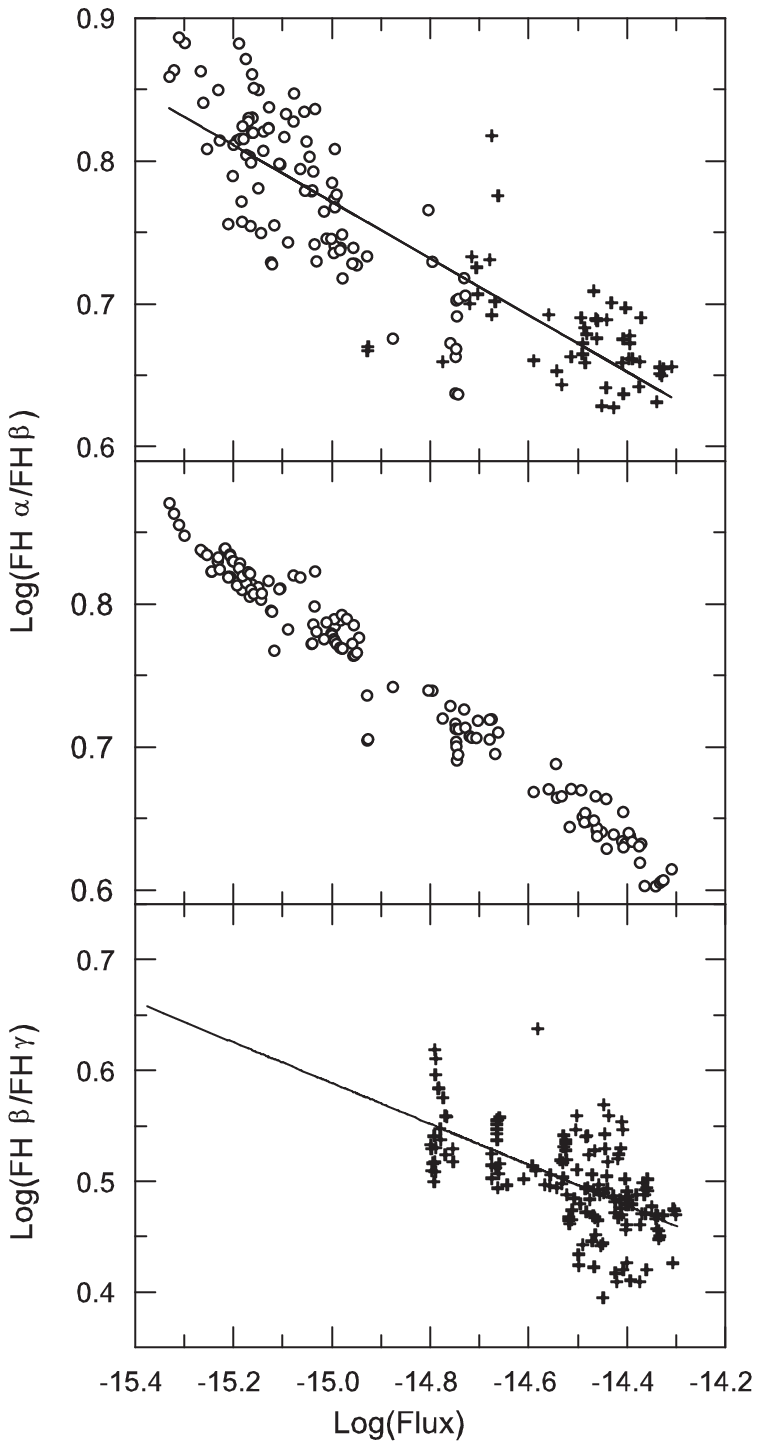}
  \caption{Balmer decrement ($F(\ha)/F(\hb)$ and $F(\hb)/F(\hg)$)
vs. optical continuum flux (\ergsA).
The scale is logarithmic on both axes.
Open circles indicate observations in
1992--2002, while crosshair symbols are the observations since 2003 when a
systematic continuum brightening begins. The solid lines is the ordinary linear
regression applied for entire period of observations.
Middle panel shows Balmer decrement vs. optical continuum flux as expected
from the Keplerian disc model (see text), while top and bottom panels show
the observed Balmer decrement.}
  \label{decrement}
\end{figure}

As was found in \citet{Ser11} the \ha\ to \hb\ ratio of fluxes
($F(\ha)/F(\hb)$) varies with the continuum flux because
the power-law index $a$ is different between \ha\ and \hb\ lines.
In Fig.~\ref{decrement} we show, in log-log scale, the $F(\ha)/F(\hb)$ (top panel)
and the $F(\hb)/F(\hg)$ (bottom panel) ratios as functions of the
optical continuum flux $F_{5100}$.
Note that the \hg\ measurements are only available since 2005.
Contributions from the host galaxy and from the narrow lines
are subtracted. The time delays  were taken into account by
the respective time shifts of the light curves. The shifted light curves
have been rebinned: the \hb\ fluxes are rebinned to the times of observations
of the \ha\ fluxes, while the \hg\ fluxes are rebinned to the times of observations
of the \hb\ fluxes.
The ordinary linear regression
gives: $F(\ha)/F(\hb) \propto F_{5100}^{-0.20}$ and
$F(\hb)/F(\hg) \propto F_{5100}^{-0.18}$. It seems that both relationships
are not power-law exactly (i.e., not linear in the log-log scale) or
there are long-term changes in these relationships.
In the middle panel of the Fig.~\ref{decrement} we show the $F(\ha)/F(\hb)$ ratio
vs. $F_{5100}$ as expected from our Keplerian disc model where the long-term
changes in the scale factors for both lines are accounted for.
The expected relationship is linear (in log-log scale)
with relatively small deviations of individual data points.
This is different from the observed relationship (top panel of Fig.~\ref{decrement}).
Probably the power-law approximation to the relationship between
Balmer line and continuum fluxes is not exactly correct.

\section[]{Summary}

\begin{enumerate}

\item
We confirm the variability of the narrow-line fluxes in the 3C~390.3 nucleus.
The \oiii\,$\lambda5007$ line flux increases monotonically by $\approx 30$ per cent
during about ten years in 2003--2014. The narrow Balmer lines show similar monotonic
increase, while variability patterns of the \oi\,$\lambda6300$ narrow line are
completely different from that of \oiii.
The observed variability of the \oiii\,$\lambda5007$ line fluxes can be
reproduced, for example, in terms of the transfer function of the spherically
symmetric shell with the inner and outer radii of 20 and 150 light-years, respectively,
and the \oiii\ emissivity per unit volume
$\varepsilon \propto r^{-3.5}$.

\item
The mean lags for the three periods of observations were found to be
$88.6\pm8.4$, $161\pm15$, and $113\pm14$\,d for the \ha, \hb, and \hg\
broad emission-lines, respectively (\hg\ line measurements are only available for the last
two periods). These values need to be corrected
for time dilation by dividing by $1+z$ to put them into the rest frame.

\item
It seems that the lags of the Balmer lines depend little,
if any, on the continuum flux.

\item
The mean reverberation mass of the central black hole in the 3C~390.3 nucleus
was found to be $(1.87\pm0.26)\times10^9\,M_\odot$ and
$(2.81\pm0.38)\times10^9\,M_\odot$, for the \hb\ and \ha\ broad
emission-lines and under $f=5.5$.

\item The black hole mass estimates from both lines are different
at $2\sigma$ confidence.
Probably, there is a difference between kinematics of the \ha\ and \hb\ emission
regions which translates to a difference in the value of the scale factor $f$
as was suggested by \citet{Ser11}.

\item The Keplerian disc model we consider in the present paper and
the model-independent calculation both show
that the reverberation mapping can only be applied to the entire
(more than twenty years) period of observations of the 3C~390.3 nucleus after
removing a long-term trend. This trend has been expressed by a variable scale
factor $c(t)$ in the power-law relationships
between line and continuum fluxes:  $F_{line}\propto c(t)\,F_{cont}^a$.
In terms of the Keplerian disc model, the origin of this trend
can be attributed to the changes in the height of the driving continuum
source above the disc centre.

\item The responses of the \ha, \hb, and \hg\ broad emission-lines
in 3C~390.3 to the optical continuum variations are non-linear.
According to our Keplerian disc model
the power-law index $a$ equals to $0.77$ and $0.54$ for the \hb\ and \ha\ lines,
respectively.
Since the power-law index $a$ is different among Balmer lines,
the Balmer decrement must vary with the continuum flux.
The observed relationship between this decrement
and the optical continuum flux is as follows:
$F(\ha)/F(\hb) \propto F_{cont}^{-0.20}$ and
$F(\hb)/F(\hg) \propto F_{cont}^{-0.18}$.
It seems that  the relationship between
Balmer line and continuum fluxes is slightly different from the
 power-law function.

\item We confirm  that the 3C~390.3 nucleus is definitely an `outsider'
in the fundamental relationship between optical luminosity
and black hole mass.  Its Eddington ratio of $E_{bol}/E_{Edd} = 0.0037$ is  very low.

\end{enumerate}

\section*{Acknowledgments}

The CrAO CCD cameras have been purchased through the US Civilian Research
and Development Foundation for the Independent States of the Former Soviet
Union (CRDF) awards UP1-2116 and UP1-2549-CR-03.

\label{lastpage}

\end{document}